\documentclass[aps,prl,twocolumn,superscriptaddress,amsfonts,amsmath,amssymb]{revtex4-1}

\usepackage{graphicx}

\usepackage{subfigure}
\usepackage{color}
\usepackage{hyperref}
\usepackage{footnote}

\begin{document}


\title{Propagation of wind-power-induced fluctuations in power grids}


\author{Hauke Haehne}
\affiliation{Institute of Physics and Forwind, Carl von Ossietzky Universit\"at Oldenburg, K\"upkersweg 70, 26129 Oldenburg, Germany}

\author{Katrin Schmietendorf}
\affiliation{Institute of Physics and Forwind, Carl von Ossietzky Universit\"at Oldenburg, K\"upkersweg 70, 26129 Oldenburg, Germany}

\author{Samyak Tamrakar}
\affiliation{Jacobs University, Department of Physics and Earth Sciences, Campus Ring 1, 28759 Bremen, Germany}

\author{Joachim Peinke}
\affiliation{Institute of Physics and Forwind, Carl von Ossietzky Universit\"at Oldenburg, K\"upkersweg 70, 26129 Oldenburg, Germany}
\affiliation{Fraunhofer IWES, K\"upkersweg 70, 26129 Oldenburg, Germany}

\author{Stefan Kettemann}
\affiliation{Jacobs University, Department of Physics and Earth Sciences, Campus Ring 1, 28759 Bremen, Germany}
\affiliation{Division of Advanced Materials Science, Pohang University of Science and Technology (POSTECH), San 31, Hyoja-dong, Nam-gu, Pohang 790-784, South Korea }


\begin{abstract}
Renewable generators perturb the electric power grid with heavily non-Gaussian and time correlated fluctuations. While changes in generated power on timescales of minutes and hours are compensated by frequency control measures, we report subsecond distribution grid frequency measurements with local non-Gaussian fluctuations which depend on the magnitude of wind power generation in the grid. Motivated by such experimental findings, we simulate the sub-second grid frequency dynamics by perturbing the power grid, as modeled by a network of phase coupled nonlinear oscillators, with synthetically generated wind power feed-in time series. We derive a linear response theory and obtain analytical results for the variance of frequency increment distributions. We find that the variance of short-term fluctuations decays, for large inertia, exponentially with distance to the feed-in node, in agreement with numerical results both for a linear chain of nodes and the German transmission grid topology. In sharp contrast, the kurtosis of frequency increments is numerically found to decay only slowly, not exponentially, in both systems, indicating that the non-Gaussian shape of frequency fluctuations persists over long ranges.
\end{abstract}


\maketitle


For the transition of electric energy supply towards renewable sources, wind power plays a crucial role. Alone in 2017, 15.6 GW of new wind power capacity was installed in the European Union (EU), now summing up to 168.7 GW, that is 18\% of the total installed power generation capacity in the EU \cite{windeurope2018}.

\begin{figure}[!h]
\centering
\includegraphics[scale=0.4]{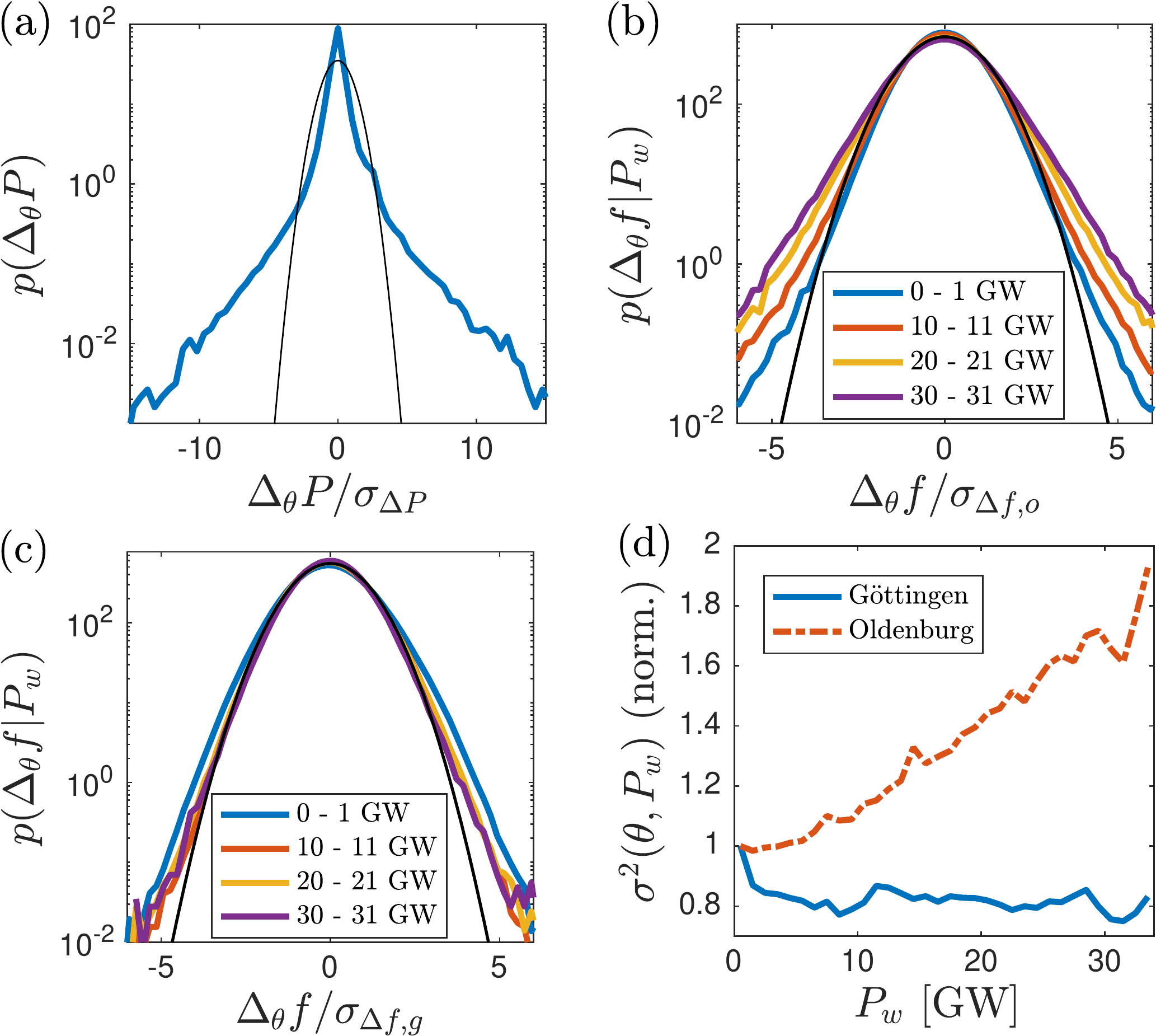}
\caption{\emph{(a)} Distribution of power increments $\Delta_\theta P = P(t+\theta) - P(t)$ of the aggregated output $P(t)$ of 12 wind turbines in a wind park (timescale $\theta = 1$ s) \citep{downloadData} and Gaussian distribution as a reference in black. Each turbine has a rated power $P_r \approx 2$ MW. Power increments are given in units of the standard deviation $\sigma_{\Delta P}= 0.023$ MW of $p(\Delta_\theta P)$. \emph{(b)} Frequency increment distribution $p(\Delta_\theta f | P_w)$ conditioned on wind energy generation $P_w$ in Germany, measured in Oldenburg ($\theta = 0.2$ s). $p(\Delta_\theta f | P_w)$ broadens with increasing $P_w$. \emph{(c)} Simultaneously measured data in G\"ottingen. Standard deviations of $p(\Delta_\theta f)$ are $\sigma_{\Delta f,o} = 0.58$ mHz in Oldenburg and $\sigma_{\Delta f,g} = 0.73$ mHz in G\"ottingen. \emph{(d)} Variance evolution of the distributions in (b) and (c) with $P_w$, normalized with the first $P_w$-bin. }
\label{fig:measurements}
\end{figure}

In contrast to the steady production of conventional power sources, wind power generation is highly volatile. Due to its continuing and increasing integration to the European power grid, maintaining a high power quality despite such fluctuations has become an important task \cite{liang2017emerging}. For the analysis of fluctuations on a timescale $\theta$ of a stochastic process $x(t)$, we study the statistics of increments $\Delta_\theta x(t) = x(t+\theta ) - x(t)$ as quantified by their moments $\langle(\Delta_\theta x - \langle\Delta_\theta x\rangle)^n\rangle$ which contain information about time correlations. Time series of wind power generation $P(t)$ show non-Gaussian increment probability density functions (PDFs) $p(\Delta_\theta P)$ on timescales of few seconds \cite{milan2013turbulent,anvari2016short}. The tails of such PDFs deviate from the Gaussian distribution even after aggregation of several turbines in a wind farm, \mbox{Fig.~\ref{fig:measurements} a}, and are related to the turbulent nature of wind speed fluctuations \cite{milan2013turbulent}.

On larger timescales, frequency control measures compensate feed-in fluctuations of renewable generators-- thereby maintaining stable grid operation \cite{operation}. However, how does the grid respond to the characteristic non-Gaussian short-term fluctuations when its dynamics is governed by the inertia of rotating masses? How does that affect power quality? Recent results from local distribution grid frequency measurements show that on timescales below 1 s, grid frequency fluctuations actually increase with increasing wind power production \cite{haehne2018footprint}. Further, the timescale separating local from interarea modes in {\it Small Signal Stability Analysis} lies in the order of 1 s \cite{zhang2012flexible}. The question arises if such wind-power-induced fluctuations are a local feature resulting from high wind power injection close to the measurement or if they rather have a long-range effect to the grid.

Dynamical systems driven by non-Gaussian, turbulence like noise model a broad range of natural and manmade phenomena \cite{friedrich2011approaching}. Up to a few seconds, the dynamics of high-voltage AC-grids are captured by networks of inert phase-coupled oscillators \cite{kundur1994power,machowski1997power,filatrella2008analysis}. For such models, the spreading of singular \cite{kettemann2016delocalization, tamrakar2018propagation}, harmonic \cite{xiaozhu}, and stochastic \cite{tyloo2018robustness} perturbations were analyzed and evaluated. Modified Fokker-Planck equations have shown to be useful to predict steady state frequency distributions obtained from measurements in highly meshed grids \cite{schafer2018non}. Further, the probability of outages caused by fluctuating perturbations was evaluated in Refs.~\cite{schmietendorf2017impact,schafer2017escape}. While these results indicate that locally induced perturbations affect the dynamics and synchronization of coupled oscillators, it is not yet understood if stochastic perturbations affect the grid only locally or if they rather propagate throughout the grid and how that depends on the parameters and intrinsic timescales of the power grid.

Our work combines experimental and theoretical analyses of short-term increments $\Delta_\theta f = f(t+\theta) - f(t)$ of the grid frequency $f(t)$ which allows us to analyze fluctuations on timescales $\theta$. We present grid frequency measurements at two different locations in Germany: Oldenburg, in the northwestern region of Germany with a high share of wind energy injection close to the measurement, and, in 213 km straight line distance towards the center of Germany, G\"ottingen, with smaller proportion of wind energy injection \cite{map}. We show that the statistics of the fluctuations depends qualitatively differently on wind power feed-in at the two measurement positions. Motivated by these observations, we study the propagation of fluctuations in power grids by performing numerical simulations of the sub second grid frequency dynamics, as modeled by a network of phase-coupled nonlinear oscillators, and derive a linear response theory to obtain analytical results for the variance of frequency increment distributions as a function of the increment timescale, the distance from the feed-in node, and the system parameters, most importantly, the inertia in the grid. \\

\emph{Data analysis.--} We use 10 kHz voltage samplings simultaneously measured in Oldenburg and G\"ottingen from 25 July 2017 until 13 March 2018, to derive a grid frequency time series $f(t)$ with a time resolution of 5 Hz. We provide details on our measurement techniques in the Supplemental Material \cite{supplement}.

We characterize the short-term frequency fluctuations in terms of width (variance) and shape (kurtosis) of their increment PDFs. We use publicly available power generation data \cite{entsoe} to obtain conditioned PDFs $p(\Delta_\theta f | P_w)$. In an earlier measurement period \cite{haehne2018footprint}, we have observed that $p(\Delta_\theta f)$ with $\theta = 0.2$ s broadens with an increased amount of wind energy $P_w$ fed to the grid in Germany. In our new data set, we were able to reproduce our earlier results in Oldenburg, Fig.~\ref{fig:measurements} b, while in the parallel G\"ottingen measurements, we see no such effect, Fig.~\ref{fig:measurements} c. For comparison, we show the evolution of the conditioned variance $\sigma^2(\theta,P_w):= \mathrm{var}(p(\Delta_\theta f | P_w))$ with $P_w$ (Fig.~\ref{fig:measurements} d) clearly confirming the different short-term response to wind power feed-in at the two measurement spots.

Both short-term increment PDFs $p(\Delta_\theta f)$, $\theta = 0.2$ s, show non-Gaussian tails, characterized by a kurtosis $k>3$, where 
\begin{align}
k = \frac{\langle (\Delta_\theta f - \langle \Delta_\theta f \rangle)^4 \rangle}{\langle (\Delta_\theta f - \langle \Delta_\theta f \rangle)^2 \rangle^2}.
\end{align}
We observe a higher kurtosis of $p(\Delta_\theta f)$ in Oldenburg ($k=4.1$) than in G\"ottingen ($k=3.4$). However, $k$ shows no clear trend with $P_w$ (not shown here).

In summary, we find that the impact of wind power generation on the short-term frequency fluctuations which is present in Oldenburg cannot be seen in G\"ottingen. Further, the PDFs $p(\Delta_\theta f)$ are slightly more heavy tailed in Oldenburg than in G\"ottingen. To explore how the width and shape of $p(\Delta_\theta f)$ evolve with distance to the volatile feed-in, we now consider a simple power grid model driven by a stochastic signal and derive an analytic expression for the variance of the increment statistics from linear response theory.\\

\emph{A network of synchronous machines.--} We consider the Synchronous Motor Model \cite{filatrella2008analysis} for high-voltage AC grids. The phase angle $\vartheta_i(t)$ is, in this model, governed by the \emph{Swing Equation}, 
\begin{align}
\tau^2\ddot{\alpha}_i &+ 2\tau\dot{\alpha}_i = \frac{J}{\gamma^2\omega_0}P_i \nonumber\\
&- \sum\limits_{j=1}^N\frac{J}{\gamma^2\omega_0}K_{ij}\sin(\vartheta_i^0 - \vartheta_j^0 + \alpha_i - \alpha_j). \label{eq:swing}
\end{align}
where $\alpha_i(t) = \vartheta_i(t) - \vartheta_i^0 $ is the deviation of $\vartheta_i(t)$ from the fixed point $\vartheta_i^0$ on a co-moving reference frame. $P_i$ denotes the generated or consumed power at node $i$. Further, $K_{ij} = K\cdot A_{ij}$, where $A_{ij}$ denotes the adjacency matrix. In the following we fix the damping $\gamma = 10^5$ kgm$^2$/s and line capacity $K = 0.5\,\mathrm{GW}$. The reference frequency is $\omega_0 = 2\pi\cdot 50\,\mathrm{Hz}$ and the internal timescale $\tau$ follows from $\tau = J/\gamma$. We use homogeneous parameters.

Let us now consider the case in which a subset of nodes $\{ j\}$ is driven by a noisy signal $P_j(t)$. The production or consumption at one of the perturbed nodes $j$ decomposes as $P_j(t) = P_j^0 + \delta P_j(t)$ into a constant value $P_j^0$ corresponding to the fixed point and a stochastic perturbation $\delta P_j(t)$ with $\langle \delta P_j(t) \rangle = 0$. 

If the system stays close to its fixed point of operation, $|\alpha_i | \ll 1$ \footnote{See Eq.~(S2) in the Supplement of \cite{tamrakar2018propagation} for a more accurate condition for the validity of linear response.}, we may calculate the response of phase $\alpha_i(t)$ to the signal $\delta\Pi_j(t):= J/(\gamma^2\omega_0)\delta P_j(t)$ in first-order approximation as
\begin{align}
\alpha_i(t) = \int\limits_{-\infty}^t \frac{dt'}{\tau}
\sum\limits_j\delta\Pi_j(t')  G_{ij} (t'-t) 
\label{eq:response-phase} 
\end{align}
where the propagator is defined by 
\begin{align} \label{eq:propagator} 
G_{ij} ( t'- t) =  \sum\limits_{n=0}^{N-1}\sum_{\sigma = \pm 1}   \frac{\phi_{ni} \phi^*_{nj}}{2 \sqrt{1-\Lambda_n}} 
(- \sigma) e^{(1 + \sigma \sqrt{1-\Lambda_n})\frac{t'-t}{\tau}},
\end{align}
see Suppl.~Eqs.~(S10-S19) for the derivation \cite{supplement}. Here, $\Lambda_n\in\mathbb{R}$ are eigenvalues and $\boldsymbol{\phi}_n\in\mathbb{R}^N$ the corresponding eigenvectors of the generalized graph Laplacian matrix $\Lambda$ with $\Lambda_{ij}=-\frac{J}{\gamma^2\omega_0}K_{ij}\cos(\vartheta^0_i-\vartheta^0_j)~~\text{and}~~\Lambda_{ii}=\frac{J}{\gamma^2\omega_0}\sum_j K_{ij}\cos(\vartheta^0_i-\vartheta^0_j)$ which is related to the stability matrix used in small signal stability analysis \cite{zhang2012flexible,milano2010power}. In contrast to prior results \cite{tyloo2018robustness, tamrakar2018propagation, auer2017stability}, our expression applies to any stochastic signal $\delta\Pi_j(t)$ and does not rely on a  pure analysis of power spectra. An expression similar to Eq.~\eqref{eq:response-phase} has been deduced in Ref.~\cite{auer2018} independently from our work.

For simplicity, we now focus on the case where the system \eqref{eq:swing} is driven at a single node $j$. With the help of Eq.~\eqref{eq:response-phase}, we derive an expression for the variance of the frequency increment PDF on timescale $\theta$, which only depends on the auto-correlation function of the increment time series $\mathrm{acf}(|\delta|) = \langle\Delta_\theta\Pi_j(t)\Delta_\theta\Pi_j(t+\delta)\rangle$,
\begin{align}
\langle\Delta_\theta\omega_i^2\rangle = \int\limits_{-\infty}^0\frac{d\tilde{t}}{\tau}\int\limits_{\tilde{t}}^{\infty} \frac{d\delta}{\tau}   \mathrm{acf}(|\delta|)\partial_t G_{ij} (\tilde{t}) \partial_t G_{ij} (\tilde{t}-\delta),
\label{eq:response-variance}
\end{align}
where $\partial_t G_{ij}$ denotes the partial derivative of the propagator $G_{ij}(t'-t)$ with respect to $t$ and $\omega_i = \dot{\alpha}_i$ the frequency at node $i$. See Suppl.~Eqs.~(S20-S27) for the derivation \cite{supplement}.\\

\emph{Localization of increment PDF variances.--} We now test our approach \eqref{eq:response-variance} numerically and analyze the propagation and localization of frequency increment statistics in linear chains of $N$ coupled oscillators obeying Swing Eq.~\eqref{eq:swing}. We emphasize that our linearization approach is principally applicable to more complex networks.

\begin{figure}[t]
\centering
\includegraphics[scale=0.68]{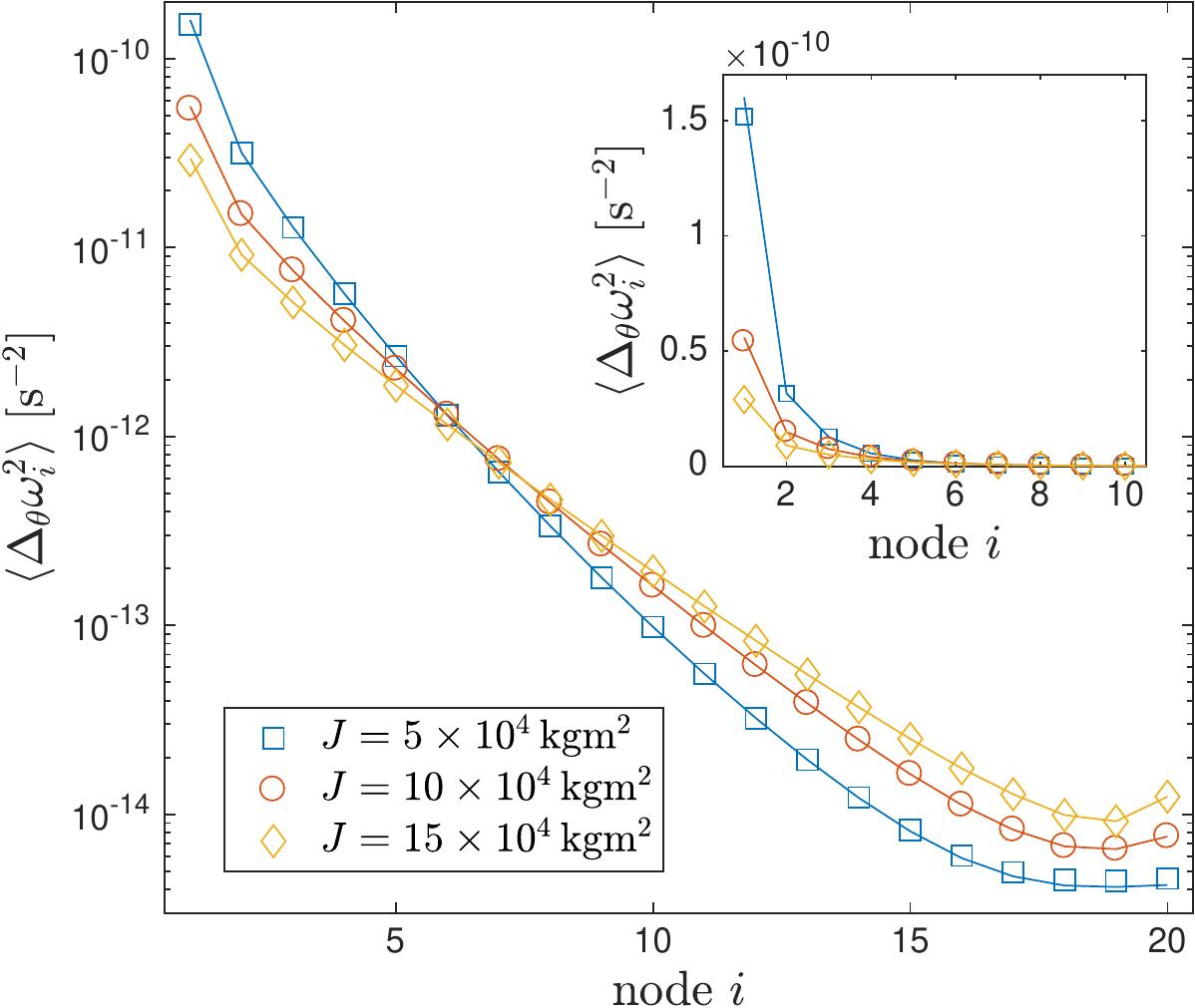}
\caption{Evolution of the variance $\langle\Delta_\theta\omega_i^2\rangle$ of increment PDFs $p(\Delta_\theta\omega_i )$ ($\theta = 0.01$ s) in a chain system of $N=20$ oscillators from linear response theory Eq.~\eqref{eq:response-variance} (solid lines) and direct Runge-Kutta simulations (symbols). Node $j=1$ is driven by the perturbation signal $\delta P(t)$. We choose $\theta = 0.01$ s because $\delta P(t)$ shows the largest deviations from the Gaussian distribution on small timescales.}
\label{fig:incs-var}
\end{figure}

In our setup, solely node $j=1$ is driven by a stochastic signal $\delta P_j(t)$, which we obtain by means of a stochastic differential
equation and subsequent modification of the power spectrum \cite{schmietendorf2017impact} (details in the Supplemental Material \cite{supplement}). 
As a result, the time series $\delta P_j(t)$ reproduces key features of wind power generation data: extreme events, 
temporal correlations, characteristic power spectrum with -5/3 decay, and heavy-tailed increment statistics. We have $\mathrm{var}(\delta P_j)=2.3$ MW and choose a grid with no initial power transfer, i.e., $P^0_i=0$ for all $i$. Increment variances obtained from direct numerical simulations of a chain of $N=20$ oscillators show good agreement with the response theory, Eq.~\eqref{eq:response-variance}, see Fig.~\ref{fig:incs-var}. The absolute amplitude of fluctuations decreases  exponentially -- apart from boundary effects -- with distance from the perturbation, confirming the localization of the fluctuations.

The impact of reduced grid inertia $J$ is of considerable importance for future power grids fed by a high share of renewable sources providing no inertia per se \cite{ulbig2014impact}. Decreasing the inertia $J$ in our model system leads, as expected, to higher fluctuation amplitudes, Fig.~\ref{fig:incs-var} inset. However, as the semi logarithmic plot reveals, decreasing $J$ while letting the damping $\gamma$ remain constant causes a \emph{faster} decay of the increment variance. So, how exactly does the exponential decay depend on the system parameters?

Increasing the inertia $J$ while keeping $\gamma$ constant increases the Laplace eigenvalues $\Lambda_n$. For chain like grids with a large number of nodes $N \gg 1$ and all nonzero $\Lambda_n >1$, we find that the variance of short-term frequency increments, Eq.~\eqref{eq:response-variance}, can be approximated by
\begin{equation}
\langle \Delta_{\theta} \omega_{i}^2\rangle \approx \frac{1}{J K \omega_0} \exp\left(- \frac{i-1}{\xi}\right) \langle \Delta_{\theta} \delta P_{1}^2\rangle, \label{eq:exp-decay}
\end{equation}
where $\langle \Delta_{\theta} \delta P_{1}^2\rangle$ is the second moment of the increment PDF of the disturbance at site $j=1$. We have used the Kolmogorov power spectrum $S(f)\propto f^{-5/3}$ of $\delta P_1 (t)$. For the derivation, see Suppl.~Eqs.~(S28-S40) \cite{supplement}. Thus, we confirm analytically that the second moment of the frequency increments is exponentially decaying with distance $i$ from the position of the disturbance and find the correlation length 
\begin{align}
\xi = v \tau/2 = \sqrt{J K}/(2 \sqrt{\omega_0} \gamma). \label{eq:corr-length}
\end{align}

\begin{figure}
\centering
\includegraphics[scale=1.1]{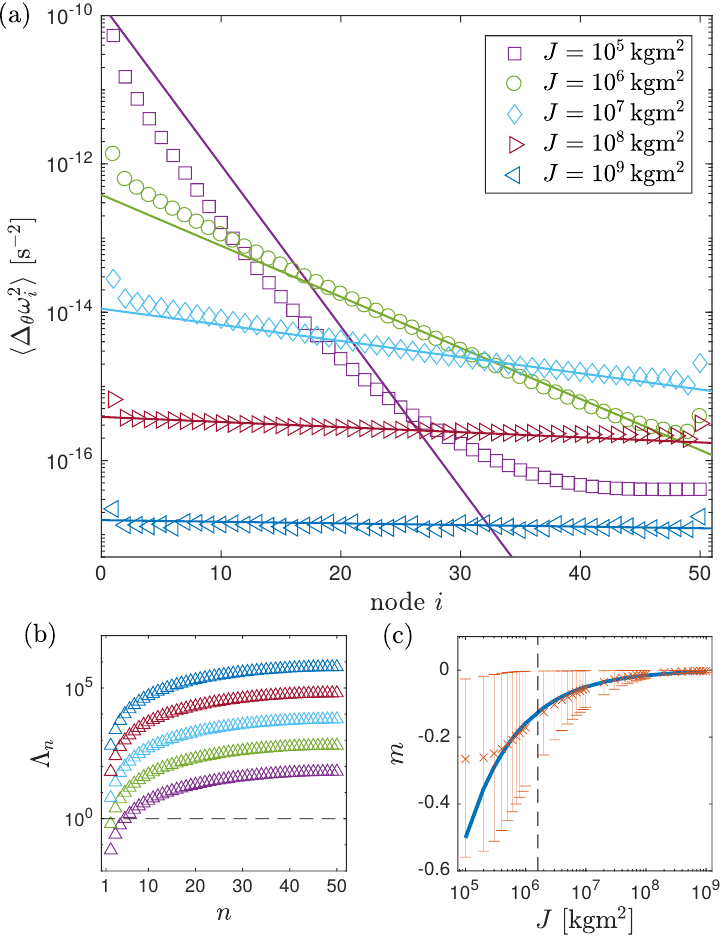}
\caption{\emph{(a)} Variance $\langle\Delta_\theta\omega_i^2\rangle$ from Runge-Kutta simulations of a chain of $N=50$ oscillators (symbols) compared to an exponential decay with slope $m=-1/\xi=-2\sqrt{\omega_0}\gamma / \sqrt{JK}$ (straight lines). \emph{(b)} Eigenvalues $\Lambda_n$ for increasing inertia $J$ (color coded, $\Lambda_n$ increase with increasing $J$). The dashed line marks $\Lambda_n =1$. \emph{(c)} Exponential fits (orange data with errorbars) of $m$ converge to our analytical prediction $m=-1/\xi$ (Eq.~\eqref{eq:corr-length}, solid blue line) for increasing inertia $J$. The vertical dashed line marks $J_c$. Error bars correspond to the 2$\sigma$ confidence bound of the exponential fit.  }
\label{fig:incs-var-estimation}
\end{figure}

We confirm the correlation length $\xi$ with numerical investigations of slightly longer chains ($N = 50$) to reduce finite-size effects, Fig.~\ref{fig:incs-var-estimation}: Once $J$ exceeds a critical value $J_c  = \omega_0\gamma^2N^2/\pi^2K \approx 1.6\cdot 10^6$ kgm$^2$, all nonzero eigenvalues $\Lambda_n$ are larger than one (Fig.~\ref{fig:incs-var-estimation} b) and the slope $m$ of the exponential decay is well fitted by $m = -1/\xi$ (Fig.~\ref{fig:incs-var-estimation} c), where $\xi$ is given by Eq.~\eqref{eq:corr-length}. In the case of low inertia $J < J_c$, there are eigenvalues with $0 < \Lambda_n < 1$. Such $\Lambda_n$ cause modes which decay more slowly with relaxation rates $\Gamma_n < 1/\tau$. If all $\Lambda_n<1$, the propagator Eq.~\eqref{eq:propagator} gets the form of a diffusion propagator \cite{kettemann2016delocalization} for which we expect that the variance of frequency increments decays with distance more slowly, with a power law. We observe the transition to such diffusive behavior in Fig.~\ref{fig:incs-var-estimation} a for $J = 10^5$ kg m$^2$, where the exponential decay only lasts up to node $i \approx 15$. Farther away, the slowly decaying soft modes dominate. \\

\begin{figure}
\centering
\includegraphics[scale=0.7]{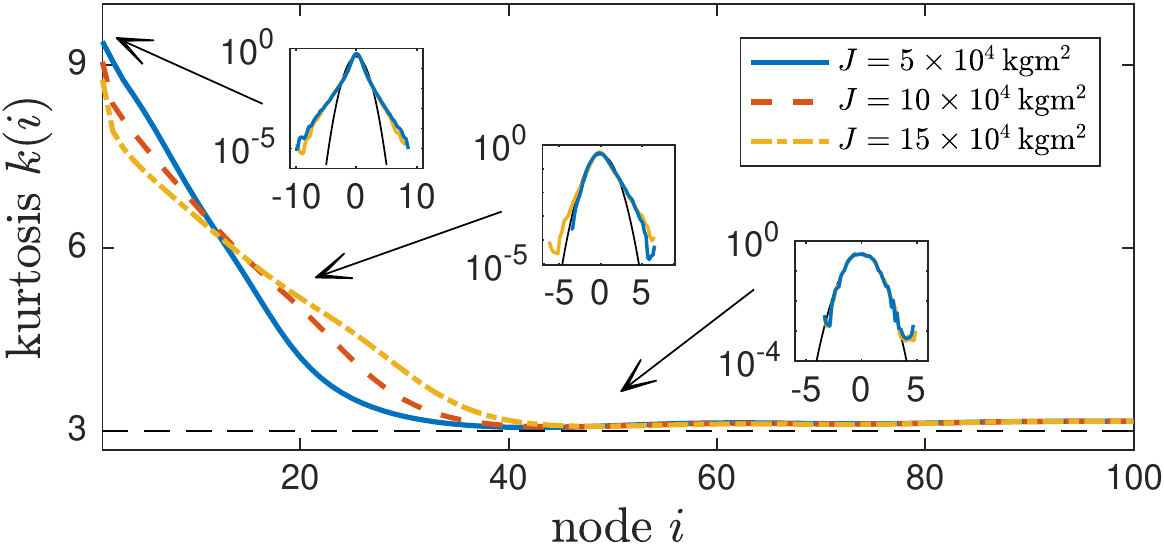}
\caption{Kurtosis $k$ of $p(\Delta_\theta \omega_i)$ in a longer chain of $N=100$ oscillators. The frequency increment distribution $p(\Delta_\theta \omega_i)$ deforms only slowly towards an (almost) Gaussian distribution (insets, from left to right $i=2,20,50$, $\theta = 0.01$ s). Results were obtained from direct Runge-Kutta simulations.}
\label{fig:incs-kurtosis}
\end{figure}

\emph{Kurtosis evolution.--} Finally, we analyze the evolution of the shape of the increment PDFs in terms of their kurtosis $k$. Interestingly, the non-Gaussian shape persists much longer than the absolute fluctuation amplitudes, see Fig.~\ref{fig:incs-kurtosis} in comparison to Fig.~\ref{fig:incs-var} inset. In numerical simulations of a chain of $N=100$ oscillators, we observe that the short-term increment PDF deforms towards an almost Gaussian distribution only after approximately 40 nodes, see also the insets of Fig.~\ref{fig:incs-kurtosis}. Surprisingly, we observe only a linear decay of the kurtosis $k$ with distance until it approaches $k \approx 3$, which corresponds to a Gaussian distribution. This slow decay is in strong contrast to the exponential decay of the fluctuation amplitude observed until approximately node 15, Fig.~\ref{fig:incs-var-estimation}, confirming that the non-Gaussian shape persists over long ranges. Again, the grid inertia seems to play a crucial role: Decreasing the inertia leads to higher kurtosis values, which means heavier tails, but also to a faster deformation towards a Gaussian distribution.\\

\begin{figure}
\centering
\includegraphics[scale=1.1]{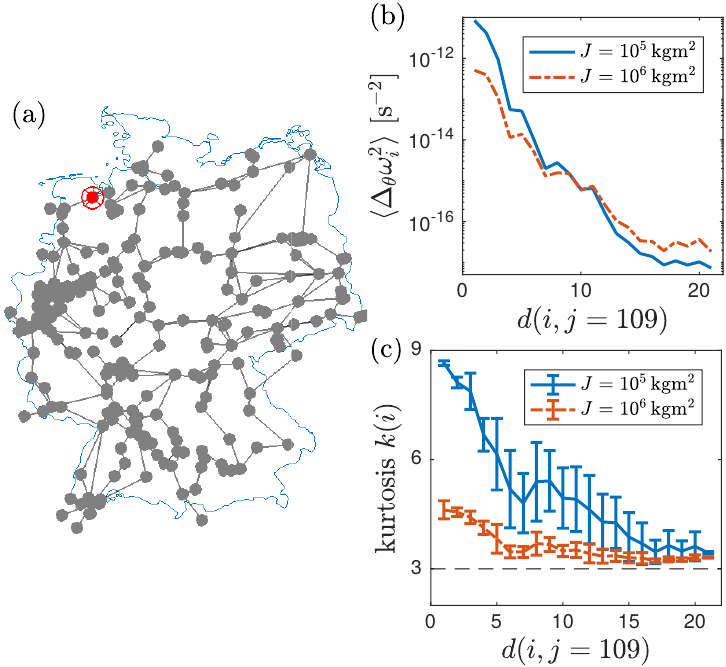}
\caption{\emph{(a)} SciGrid topology of the 380 kV grid in Germany \cite{matke2016scigrid} with homogeneous parameters as in the preceding simulations. The fluctuating perturbation $\delta P_i(t) $ is injected at node $j=109$ (highlighted in red). \emph{(b)} Variance of the short-term increments $\Delta_\theta\omega_i$ plotted against distance $d(i,j)$ to the perturbation. Here, we choose the shortest path distance and average the variances for all nodes with the same $d(i,j)$. The timescale is $\theta = 0.01$ s. \emph{(c)} Evolution of the kurtosis $k(i)$ of the increments $\Delta_\theta\omega_i$. The errorbars correspond to the standard deviation of the kurtosis of nodes with the same distance $d(i,j)$. In (b), we omit errorbars due to the logarithmic $y$-axis.}
\label{fig:scigrid}
\end{figure}

\emph{Complex network.--} In a numerical study, we show that the propagation of fluctuations is qualitatively similar on a complex topology, Fig.~\ref{fig:scigrid}. We choose the SciGrid topology \cite{matke2016scigrid} which we perturb at a single node, Fig.~\ref{fig:scigrid} a. The decay of the variance of short term increments $\Delta_\theta\omega_i$ is getting steeper with decreasing inertia $J$ and approximately follows an exponential curve, Fig.~\ref{fig:scigrid} b. Similar to our observation in linear chains, the non-Gaussian shape persists over long ranges: The kurtosis values are larger for low inertia and decrease slowly, not exponentially, with distance to the perturbation, Fig.~\ref{fig:scigrid} c.\\

\emph{Conclusion.--} In this article, we have analyzed the spreading of short-term frequency fluctuations in power grids induced by wind power feed-in. While the influence of wind power is clearly visible in the short-term fluctuations in Oldenburg (with a large installed capacity of wind power generators \cite{map}), the G\"ottingen measurements (with a rather small installed capacity \cite{map}) do not show such an effect. Our analytical and numerical investigations show that the amplitudes of such short-term fluctuations damp out exponentially fast and form a local stochastic property of the frequency-- as expected from our measurement. In contrast, the non-Gaussian shape of the frequency increment PDFs persist much wider in the grid and decays, in terms of kurtosis, only linearly with distance to the perturbation. Effects of topology and heterogeneity could further contribute to the effect we observe in the data, and that will be a subject of further research. Here, we focused on the most basic mechanism, namely, if the spatial propagation can explain our observation.

The amplitudes of the frequency fluctuations we observe are small (few mHz, Fig.~\ref{fig:measurements}). Hence, they do not cause risks for outages. Short-term fluctuations are expected to further decrease when the feed-in of many wind farms is aggregated. However, in future power grids mainly fed by wind and solar power, the interplay between the locality of short-term fluctuations and wide-scale averaging will be of interest for maintaining a high power quality. Further, we emphasize the subtle role of long-ranging soft modes induced by reducing the inertia in the grid.

Our analytical expressions help to estimate the timescale-dependent impact of fluctuations on power grids: For short timescales, this concerns, for example, the configuration of power converters feeding wind power to the grid. However, our analysis could be extended to longer timescales of minutes when the dynamics of primary and secondary control are included. On such timescales, the amplitudes of power increments $\Delta_\theta P$ of wind turbines are much larger and, hence, our theory may then help to develop new strategies for decentral frequency control. Further, our analysis of small-signal fluctuations in power grid frequency measurements, well situated within the linear response regime, will help to elaborate general results on networks of coupled oscillators, such as transient spreading dynamics \cite{wolter2018quantifying} or optimal noise-canceling topologies \cite{ronellenfitsch2018optimal}.\\
 
\begin{acknowledgments}
We thank M. Benderoth and M. Timme for enabling us to measure the grid frequency at MPI-DS in G\"ottingen and the electronic workshop at Carl von Ossietzky University Oldenburg, especially T. Madena, for technical support. We acknowledge support from the ministry for science and culture of the German federal state of Lower Saxony (grant No. ZN3045, nieders. Vorab) to H.H. and from BMBF (CoNDyNet, FK. 03SF0472D) to S.K.
\end{acknowledgments}


%

\end{document}